\def\year{2021}\relax
\documentclass[letterpaper]{article} 
\usepackage{aaai21}  
\usepackage{times}  
\usepackage{helvet} 
\usepackage{courier}  
\usepackage[hyphens]{url}  
\usepackage{graphicx} 
\usepackage{natbib}  
\usepackage{caption} 
\title{Preventing Repeated Real World AI Failures by Cataloging Incidents: \\The AI Incident Database }
\author{

    Sean McGregor
    \\
}
\affiliations{

    XPRIZE Foundation, Partnership on AI,\thanks{Representing the XPRIZE Foundation as a Partnership on AI non-profit partner.} Syntiant Corp.\\

    7555 Irvine Center Dr.
    Suite 200\\
    Irvine, California 92618\\
    iaai21@seanbmcgregor.com

}

\begin{document}

\maketitle

\begin{abstract}
Mature industrial sectors (e.g., aviation) collect their real world failures in incident databases to inform safety improvements. Intelligent systems currently cause real world harms without a collective memory of their failings. As a result, companies repeatedly make the same mistakes in the design, development, and deployment of intelligent systems. A collection of intelligent system failures experienced in the real world (i.e., incidents) is needed to ensure intelligent systems benefit people and society.
The AI Incident Database is an incident collection initiated by an industrial/non-profit cooperative to enable AI incident avoidance and mitigation. The database supports a variety of research and development use cases with faceted and full text
search on more than 1,000 incident reports archived to date.
\end{abstract}

\section{Introduction}

Governments, corporations, and individuals are increasingly deploying intelligent systems to safety-critical problem areas, including transportation \cite{NTSB2017} and law enforcement \cite{Dressel2018}, as well as challenging social system domains such as recruiting \cite{Dastin2018}. Failures of these systems pose serious risks to life and wellbeing, but even good-intentioned intelligent system developers fail to imagine what can go wrong when their systems are deployed in the real world. Worse, the artificial intelligence system community has no formal systems whereby practitioners can discover and learn from the mistakes of the past. Individuals in technology \cite{Olsson2019}, legal practice \cite{Hall2020}, and reputation management \cite{Pownall2020} now collect artificial intelligence failure history on Google Docs and GitHub. While these are admirable efforts, a person checking for problems matching their technology or problem domain will need to page through lists of links to find ones of potential relevance. Existing lists are difficult to use in development, are not comprehensive archives, and are representative of individual viewpoints of artificial intelligence (AI) failures in the real world.

Avoiding repeated AI failures requires making past failures known to AI practitioners. Therefore, we introduce a systematized collection of incidents where intelligent systems have caused safety, fairness, or other real world problems. The AI Incident Database (AIID) answers the question, ``what can go wrong when someone deploys this system"?

The contributions of this work are 3 fold. We provide infrastructure supporting best practices within the artificial intelligence industry, a dataset of more than 1 thousand incident reports, and an architecture for building research products on the growing collection of incidents.
We begin by exploring incident databases in other fields of practice before introducing the system architecture of the AIID. We then wrap up with a few concluding remarks.

\begin{figure}[t]
\centering
\includegraphics[width=0.9\columnwidth]{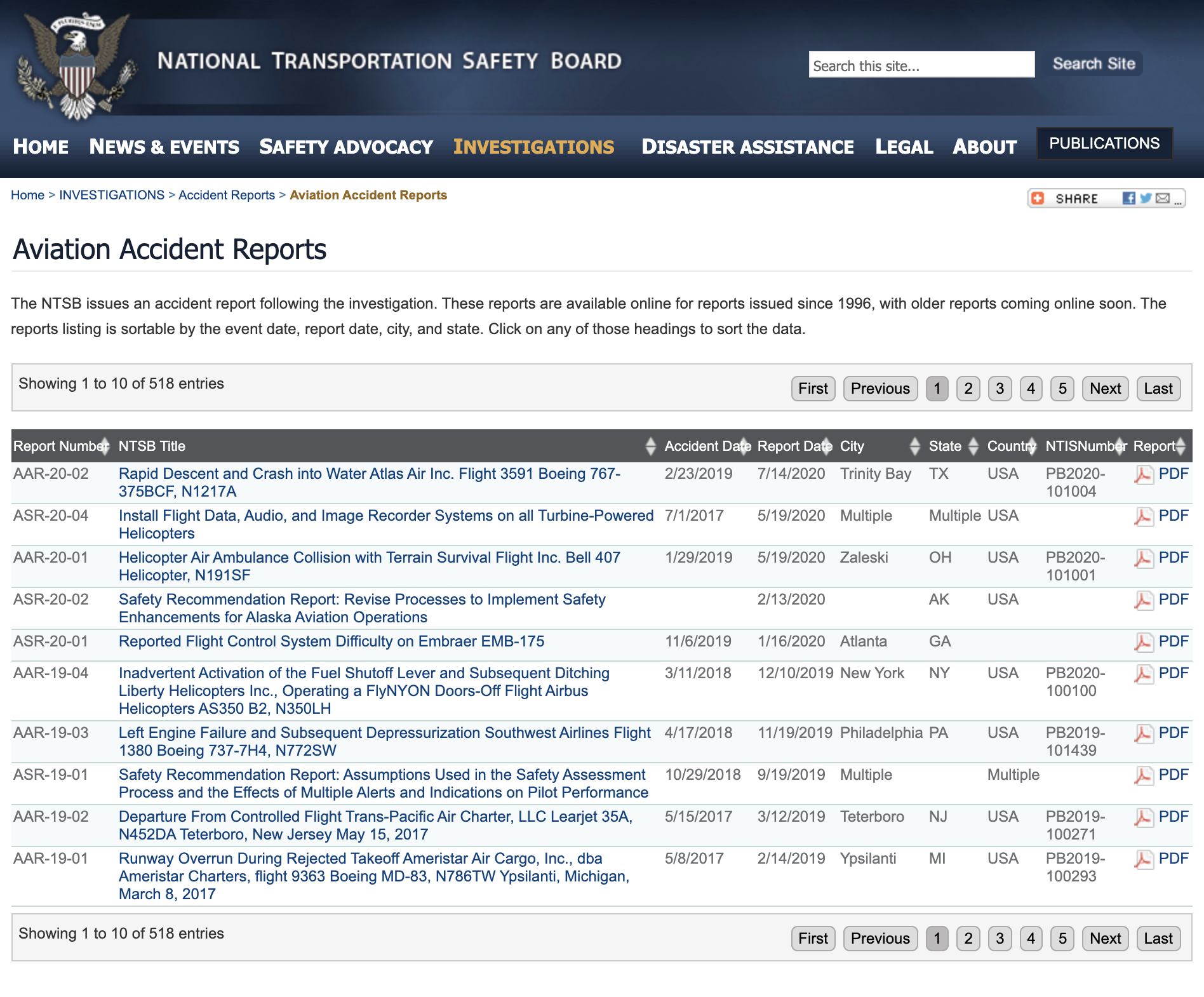}
\caption{
The US National Transportation Safety Board's (NTSB's) database shown above indexes incident and accident timelines, location, meteorology, severity, aircraft, operators, and phase of flight as facets. The reports also have a full text narrative that is searchable. Upon completion of an investigation, the report is indexed alongside the case record within the database \cite{FederalAviationAdministration2020}.
}
\label{fig:FAA}
\end{figure}

\section{Other Incident Databases}

The commercial air travel industry owes much of its increasing safety to systematically analyzing and archiving past accidents and incidents within a shared database. In aviation, an accident is a case where substantial damage or loss of life occurs. Incidents are cases where the risk of an accident substantially increases. For example, when a small fire is quickly extinguished in a cockpit it is an ``incident" but if the fire burns crew members in the course of being extinguished it is an ``accident." The aviation database (see Figure \ref{fig:FAA}) indexes flight log data and subsequent expert investigations into comprehensive examinations of both technological and human factors. In part due to this continual self-examination, air travel is one of the safest forms of travel. Decades of iterative improvements to safety systems and training have decreased fatalities 81 fold since 1970 when normalized for passenger miles \cite{Mediavilla2019}.

Aviation accidents share a well-defined operational context, but intelligent systems can be applied to all contexts. The comprehensive nature of ``intelligence" means AI incident databases ingest unforeseen and novel contexts, technologies, and failures. The AIID design outlined in the next section introduces a system architecture inspired by the aviation incident and accident database, but with a greater emphasis on extensibility.

The second incident database inspiring the AIID is the Common Vulnerabilities and Exposures (CVE) system, which contains $141,076$ publicly disclosed cybersecurity vulnerabilities and exposures \cite{TheMITRECorporation2020}. In contrast to the aviation database, which serves users associated with a single industry, the CVE site serves as critical security infrastructure across all industries by enabling vulnerabilities to be circulated and referenced with a consistent identifier. Other systems consume the identifiers to apply taxonomies (e.g., the Common Vulnerability Scoring System), produce research, and develop more secure software. 
The creation of numbered identifications forms community infrastructure that the field of artificial intelligence currently lacks. The lists of \citet{Olsson2019}, \citet{Hall2020}, and \citet{Pownall2020} lack the comprehensive coverage, identification, and extensibility properties of the CVE, and the full text search capability of the NTSB database.

\section{The AI Incident Database}

\begin{figure}[t]
\centering
\includegraphics[width=0.8\columnwidth]{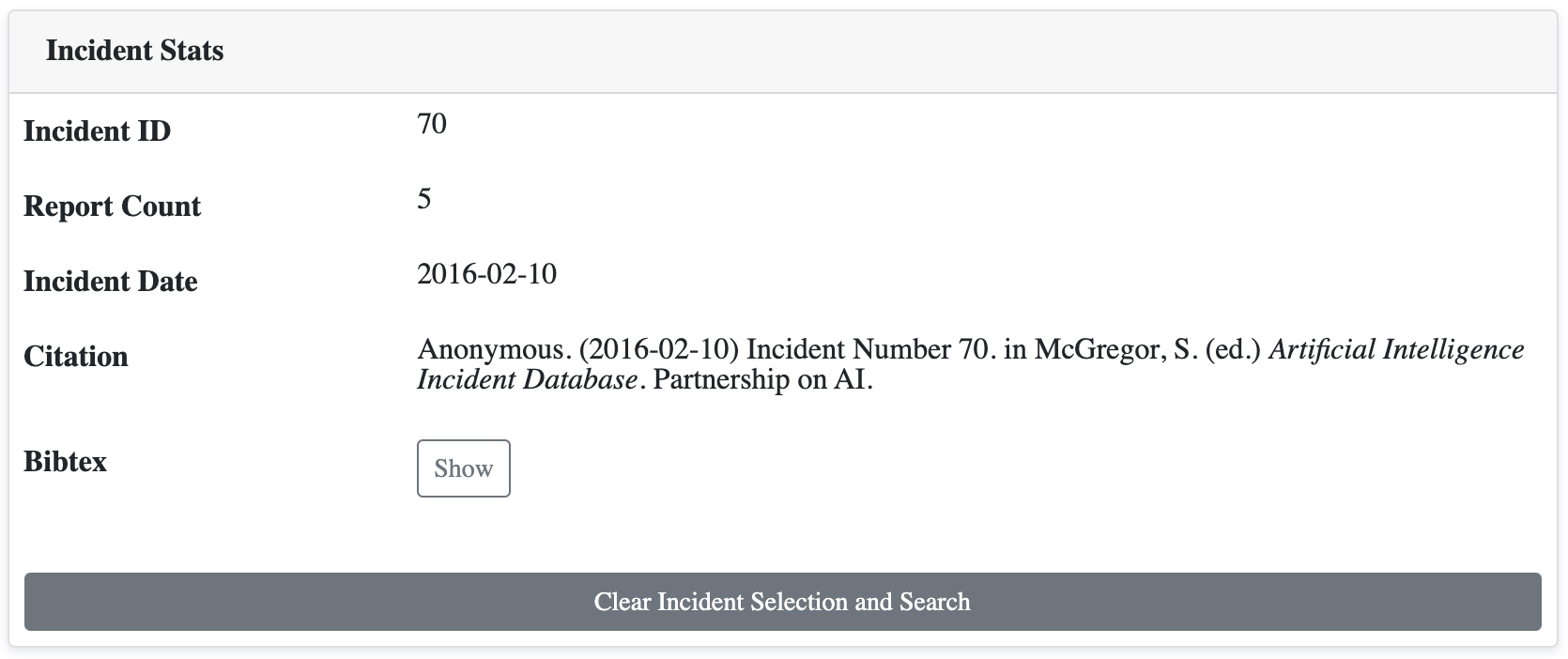}
\caption{Incident numbers enumerate AI incidents within the database and join together incident reports. Incidents are citeable and are credited to the person who first submitted an incident report for a new incident. The incidents are intentionally untitled because incident descriptions tend to change through time as more is learned and achieving consensus titles by both corporate and non-profit database participants would be difficult.}
\label{fig:incidentstats}
\end{figure}

The AIID indexes more than 1,000 publicly available ``incident reports," which are a mixture of documents from the popular, trade, and academic press. Multiple reports often pertain to a single incident collectively joined together by a single identifier. For example, incident number 3 is composed of 18 reports on the Boeing 737 MAX 8 crashes \cite{aiid:3}. The variety of reports serves several purposes. First, it provides multiple viewpoints on incidents for which there is often disagreement about fair characterizations. In the Boeing case, people disagree on the extent to which technological or human factors played a part in the tragedies. Second, the number of publications and publication types serves as a proxy for interest in the incident. More reported incidents are typically more damaging, more sensational, or both. After opening the AIID to public submissions, we expect incident 3 will have thousands of incident reports due to intense public interest in the safety of flight. Lastly, sampling multiple reports per incident gives a more complete coverage of the words applicable to an incident and increases the likelihood of users discovering incidents relevant to their use cases. The use cases are detailed in the following user stories.

\textbf{User: Product Managers}.
Corporate product managers are responsible for defining product requirements before and during product development. If a product manager discovers incidents where intelligent systems have caused harms in the past, they can introduce product requirements to mitigate risk of recurrence. For example, when a product manager is specifying a recommender system for children, the AIID should facilitate the discovery of incident 1 \cite{aiid:1}, wherein YouTube Kids recommended inappropriate content. Knowledge of incident 1 would produce a range of technological, marketing, and content moderation requirements for the product.

\textbf{User: Risk Officers}.
Organizationally, risk officers are tasked with reducing the strategic, reputational, operational, financial, and compliance risks associated with an enterprise's operation. Consider the case of a social network preparing to launch a new automatic translation feature. A search of ``translate" within the AIID returns 40 separate reports, included among them an incident wherein a social media status update of ``good morning" translated to ``attack them" and resulted in the user's arrest \cite{aiid:72}. After discovering the incident, the risk officer can read reports and analyses to learn that it is currently impossible to technologically prevent this sort of mistake from happening, but there is a variety of best practices in mitigating risk, such as clearly indicating the text is a machine translation.

\textbf{User: Engineers}.
Engineers can also benefit from checking the AIID to learn more about the real world their systems are deployed within. Consider the case of an engineer who is making a self driving car with a image recognition system. The experience of incident 36 \cite{aiid:36}, where a woman in China was shamed for jaywalking because her picture was on the side of a bus, shows hows images can confuse image recognition systems. Such cases must therefore be represented within safety tests.

\textbf{User: Researchers}.
Safety and fairness researchers already employ case study methodologies in their scholarship, but they presently lack the capacity to track AI incidents at the population level. For example, it is difficult to show the rate at which incidents involving policing are changing through time. An AIID search for ``policing" in the full text of reports currently returns 14 distinct incidents. Each of these incidents are additionally citeable (see Figure \ref{fig:incidentstats}) within research papers. The resulting research papers can then be added to the database as further reporting on the incident. Additionally, researchers can show the importance of their publications by citing incidents that could potentially be mitigated through their advances.

Finally, we note that making a database entry shareable (i.e., linkable) empowers these users rhetorically to convince others that mitigation is necessary. Technology companies are famous for their penchant to move quickly without evaluating all potential bad outcomes. When bad outcomes are enumerated and shared, it becomes impossible to proceed in ignorance of harms.

\subsection{System Architecture}

\begin{figure*}[t]
\centering
\includegraphics[width=1.8\columnwidth]{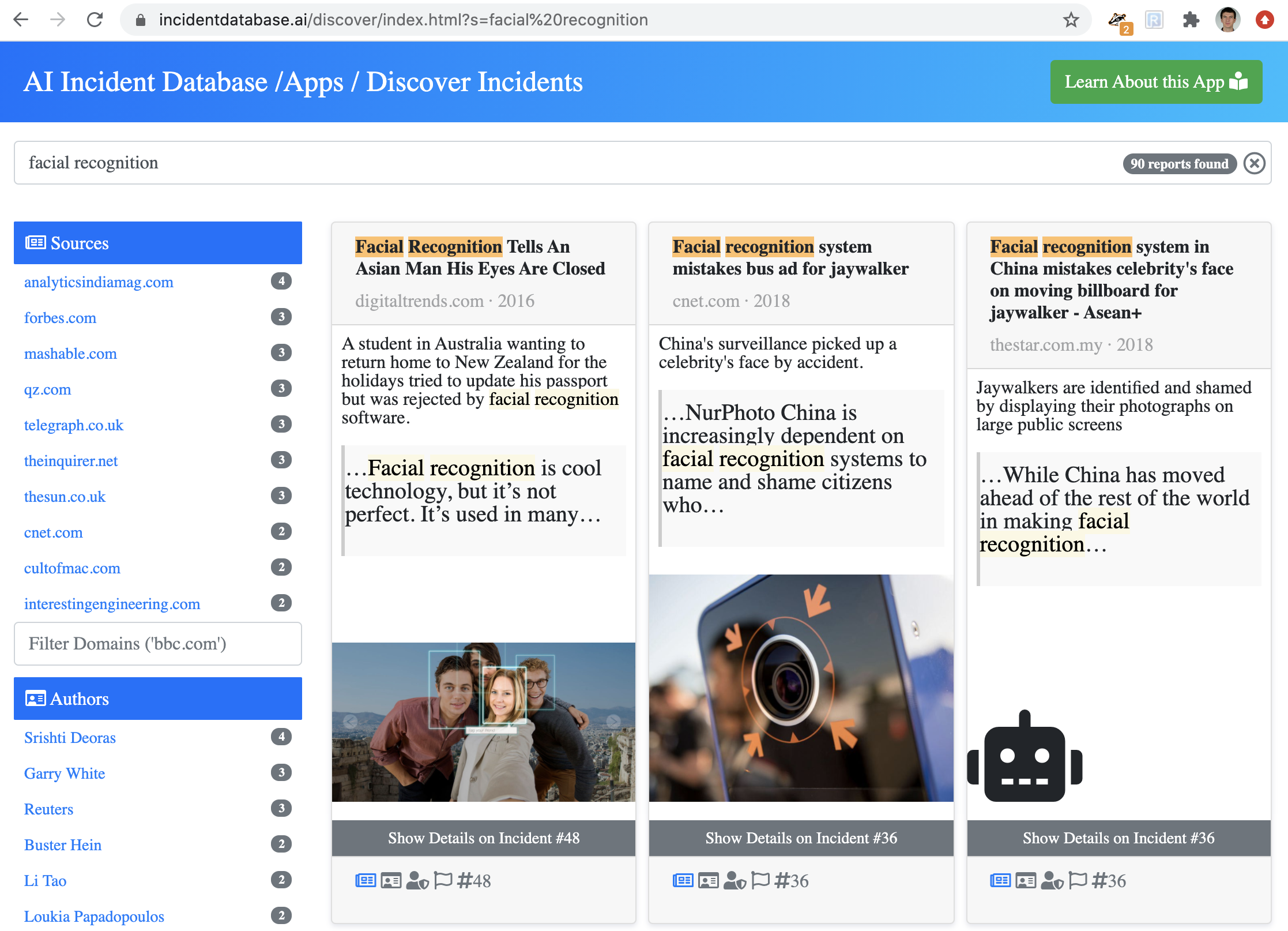}
\caption{A user has entered ``facial recognition" as a search term into the search box of the ``Discover" AIID application. 89 reports returned to the search instantaneously (every keystroke filters the results and the page renders) and the matching text from the reports is snippeted. The publications represented within the results are faceted in the left column along with the authors, submitters, and incident numbers to support filtering the reports based on their metadata.}
\label{fig:facialreco}
\end{figure*}

The AIID is a project of the Partnership on AI (PAI), which is a multi-stakeholder organization funded by technology companies and governed by a board of directors split between corporate partners and non-profit civil society organizations. The dual nature of the organization means prescriptive norms are difficult to promulgate. For example, in 2018, Google unveiled a dataset and competition meant to combat bias in computer vision classifiers \cite{Doshi2018}. Discussion within the partnership's Fairness, Transparency, and Accountability working group briefly entertained providing a PAI endorsement for the program, but such an endorsement proved impossible due to vigorous debate on whether computer vision should be improved in human contexts. Such debates are both a great weakness of PAI and also its greatest asset as a multi-stakeholder organization. Through forced exposure to diametrically opposed views, it becomes possible for both sides to incrementally iterate and find common ground. The AIID provides infrastructure joining together these disparate viewpoints so that they can speak for themselves and avoids providing top-down analysis or narrative. Instead, it is the responsibility of the open source community to layer taxonomies and summaries onto the database. This mirrors practices by the CVE system, which provides the identifier to a surrounding security community ecosystem. 

\begin{figure}[t]
\centering
\includegraphics[width=0.9\columnwidth]{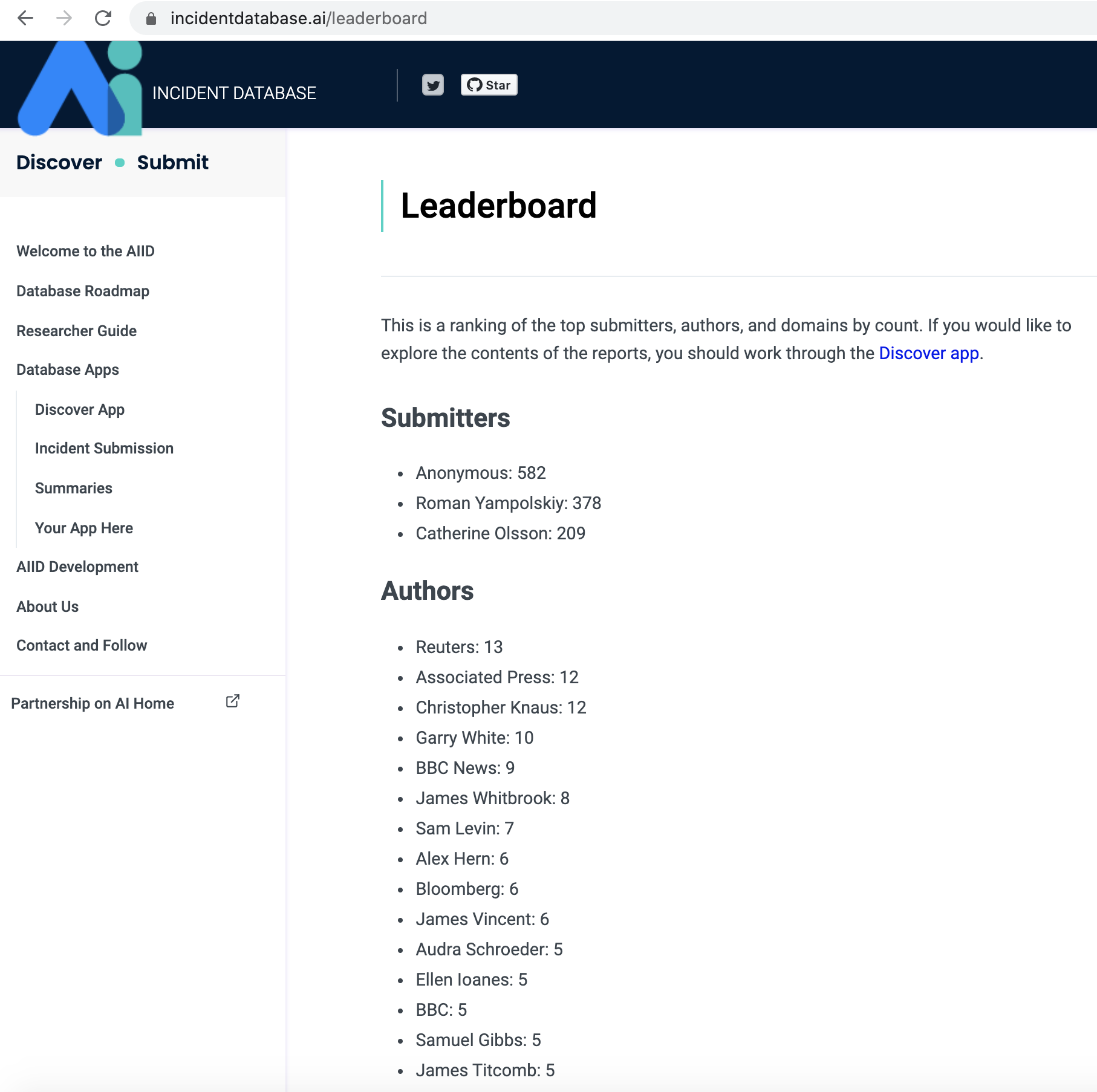}
\caption{The database provides a leaderboard of submitters and authors totalling the number of reports associated with their submissions. Gamification in other contexts has shown that people are more eager to volunteer their time for a community resource if the sum total of their contribution is constantly recognized and reinforced.}
\label{fig:leaderboard}
\end{figure}

\begin{figure}[t]
\centering
\includegraphics[width=0.9\columnwidth]{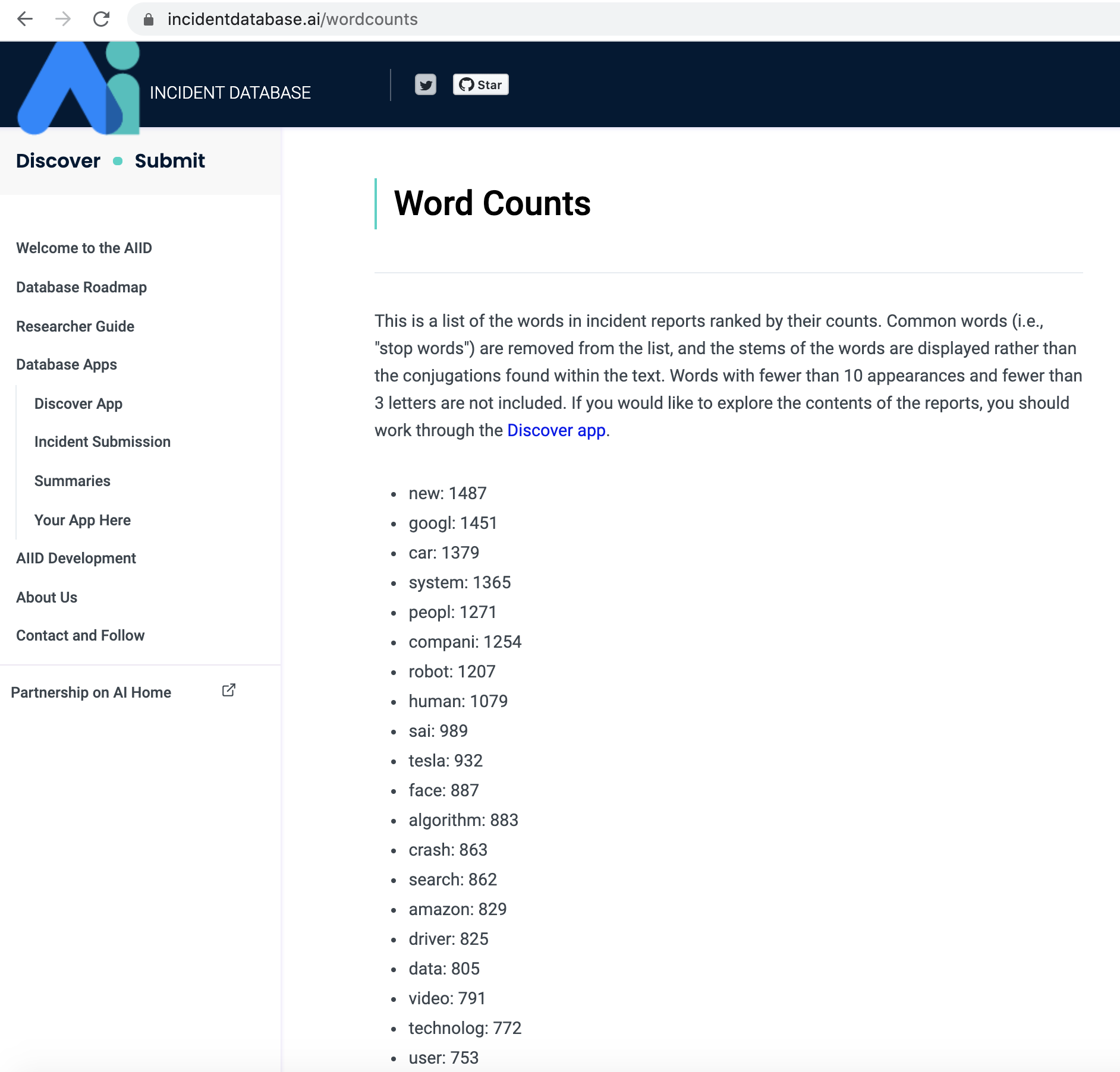}
\caption{Every time the applications rebuild, the complete text of all reports are queried, stemmed, and stop words are removed. The top words in the database are then rank ordered and rendered in the page. This application generates no requests to the database when a user requests it since the page is pre-built in the application rendering pipeline. This means computationally expensive natural language models could be applied in the application lifecycle (e.g., training topic models) without negatively impacting user experience.}
\label{fig:wordcounts}
\end{figure}

The AIID is a collection of web applications that interfaces with a MongoDB document database storing incident report text and metadata. The first application developed for the database is the ``Discover" application, which is built to help users discover past incidents relevant to their work. Figure \ref{fig:facialreco} shows one search in the Discover application. All searches in the Discover application are ``instant searches," meaning they return results in less than a second. Another application is the ``Submit" application, which is a form for submitting links to publicly available incident reports. As the database grows, the Submit application will grow in sophistication to resolve new report submissions to existing incidents. Incident reports are often written and submitted long after the initial creation of an incident record and resolving new submissions to already indexed incidents will require additional tooling.

\begin{figure}[t]
\centering
\includegraphics[width=0.5\columnwidth]{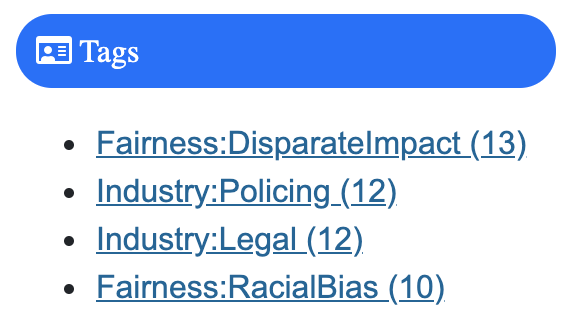}
\caption{A mockup of two taxonomies presented as facets within the Discover application. Here the Fairness tags could be defined and applied by a non-profit active in fairness advocacy, while the Industry tags could be defined and managed by a business consultancy. While the tags are all controlled by their own application and managing entity, they can be applied as filters across all applications within the AIID.}
\label{fig:taxa}
\end{figure}

\begin{figure*}[t]
\centering
\includegraphics[width=1.9\columnwidth]{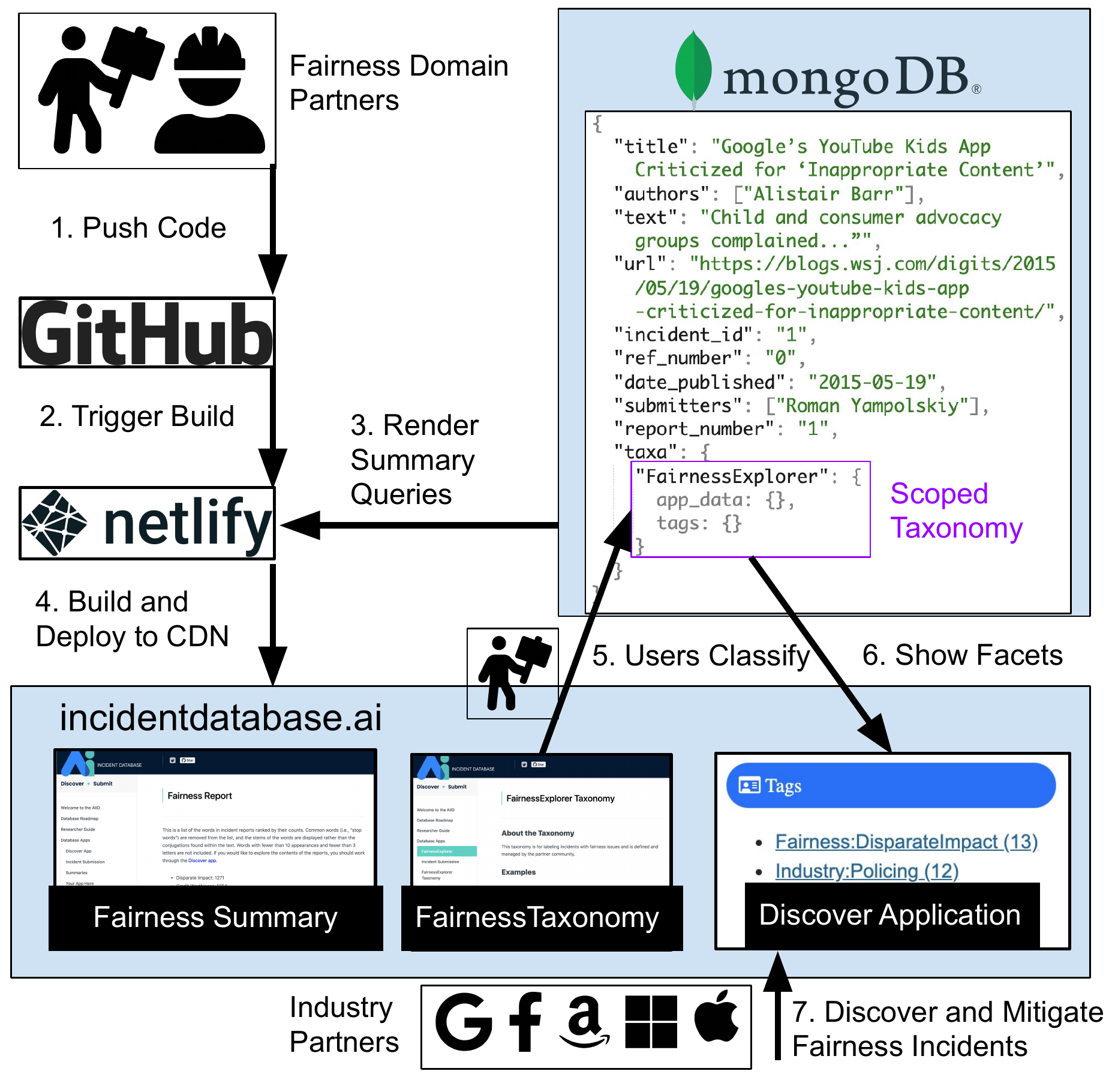}
\caption{Fairness domain partners create a new application for the AIID named ``FairnessExplorer," which includes both a taxonomy and a summary page. Upon pushing the code to GitHub, a build on the static website hosting service Netlify is triggered. The build process queries the database to generate static summaries of the database contents, including the Fairness Summary. The website then deploys to a global content distribution network. Users can then apply classifications within the scope of the FairnessExplorer taxonomy. Industry partners can then filter incidents based on the classifications of the FairnessExplorer taxonomy, as well as taxonomies developed by other domain partners.}
\label{fig:system_architecture}
\end{figure*}

The system architecture is built to express the full range of viewpoints represented by the PAI partner community by allowing partners to build their own taxonomies, taxonomy documentation, and data summaries (see Figure \ref{fig:system_architecture}). The taxonomy system is built to be interoperable, meaning tags are namespaced and can be faceted (i.e. filtered). For example, Figure \ref{fig:taxa} shows how taxonomies defined in Industry and Fairness namespaces can be faceted within the Discover application. Namespaces are managed by individual or groups of partners and thus are not necessarily representative of the whole PAI ecosystem. This avoids the challenge of developing a single shared universal ontology for AI incidents and instead allows for multiple viewpoints on the data to develop and compete for mindshare.
All incident reports have metadata captured on entry into the database, including title, source, author, submitter, publication date, incident date, and incident number. These are all objective facts that can be filtered as shown in Figure \ref{fig:facialreco}. These facets are also where we introduce subjective taxonomic classification of reports and incidents.

The applications are hosted within the context of a web application combining documentation, data products (see Figure \ref{fig:wordcounts}), and social credit (see Figure \ref{fig:leaderboard}). The problem with these database views is that they often require iterating over the complete database. If these pages render for the user every time the user visits the page, the database would be slow and expensive to host. Instead, the AIID periodically pre-renders database views as static web applications, which means they only require a single database request at the time the website builds. As such, it is possible to develop a gallery of views into the data similar to the D3JS gallery, which has 168 different visualization examples \cite{Bostock2020}. One potential AIID example is supporting trend analysis for unsupervised topic models or hand tailored topic models monitoring technology, affected populations, or applications through time. These analyses can be incorporated into the static build (see Figure \ref{fig:system_architecture}) and update automatically when the website updates.

\section{Conclusion}

We expect the extensible architecture will provide for the most pragmatic coverage of AI incidents through time while reducing negative consequences from AI in the real world. Early indications of adoption are strong. Even prior to publishing the database, we have received collaboration requests from ``Big 4" accounting firms, international consultancies, law firms, research institutes, and individual academics. Through time we hope the database will develop from the work product of a small team of individuals into community owned infrastructure aligned with producing the most beneficial intelligent systems for people and society. To quote Santayana, ``Progress, far from consisting in change, depends on retentiveness... Those who cannot remember the past are condemned to repeat it." \cite{santayana1924life}

\section{Acknowledgments}
The author is grateful to Jingying Yang and the ABOUT ML project at the Partnership on AI for sponsoring this work and to the incident submitters, especially to Roman Yampolskiy, Catherine Olsson, Sam Yoon, Patrick Hall, and Charlie Pownall for their efforts in collecting many incidents to date. Iftekhar Ahmed also provided useful feedback on the paper.

\bibliography{citations}
\end{document}